%% file: WCh_060922.tex
\documentclass[12pt,a4paper]{article}
\usepackage{amsfonts}
\usepackage{mathrsfs}  
\usepackage{amssymb}

\input makra

\begin{document}
\title{Ground state energies of the Hubbard models and the
Hartree Fock approximation}
\author{\em Jacek Wojtkiewicz {\rm and} Piotr H. Chankowski,\\
Faculty of Physics, University of Warsaw,\\
Pasteura 5, 02-093 Warszawa, Poland\\
\footnotesize{e-mails:} ${\rm wjacek@fuw.edu.pl}$ (J.W.),
${\rm chank@fuw.edu.pl}$ (P.H.Ch.)
}
\maketitle
\abstract{
  According to the `folk knowledge', the Hartree-Fock (H-F) approximation
  applied to the Hubbard model becomes exact in the limit of small coupling
  $U$ (the smaller $|U|$, the better is the H-F approximation). In \cite{BP}
  Bach and Poelchau have substantiated a certain version of this assertion
  by providing a rigorous estimate of the difference between the true
  ground-state energy of the simplest version of the Hubbard model and the
  H-F approximation to this quantity. In this paper we
  extend their result in two directions: i) we relax the assumption
  about the strict translational invariance of the hopping matrix,
  ii) we prove an analogous estimate for a class of multiband Hubbard
  models.}
\vskip0.3cm


\section{Introduction}
\label{sec:Intro}

The Hubbard model has been introduced almost 60 years ago
\cite{Hubbard,Gutzwiller,Kanamori}, originally to explain ferromagnetism
of transition metals. Although this attempt was not successful, the model
and its multiband extensions play, since then, the role of a `minimal
models' for strongly correlated electron systems and is broadly
used to model phenomena, which cannot be explained within the simple
one-particle picture, provided e.g. by the band theory. It is believed that
the Hubbard model (and/or its multiband extensions) qualitatively captures
the essence of such phenomena as a variety of magnetic orderings (ferro-
and antiferromagnetic as well as more complicated magnetic structures),
metal--insulator transitions and the high-temperature superconductivity.
More recently the fermion and boson versions of the Hubbard model have
been successfully applied to model atomic gases in traps.

However, despite being only a simplified model of complex real physical
systems, the Hubbard model is `notoriously difficult' (Lieb, \cite{Lieb}):
exact solutions or rigorous results are rare even in the case of its
simplest one-band variant. The same can be also said of reliable
approximations to its solutions. The most frequently used one is the
Hartree-Fock (H-F) approximation which introduces substantial
simplifications but remains still non-trivial (see for instance \cite{BLS}
and \cite{BLTrav}). There is a broad consensus that this approximation
works well for small values of the Coulomb coupling constant $U$ of the
interaction term of the Hubbard model Hamiltonian. An important step
towards justifying this expectation has been made by Bach and Poelchau
\cite{BP} (see also \cite{Bach_ICMP}) who have rigorously proved an
inequality which provides an estimate of the difference between the
true ground-state energy of the simplest version of the Hubbard model
and the H-F approximation to this quantity.

The crucial element of the derivation of the estimate given in \cite{BP}
is relating a certain constant appearing in one of the established
inequalities to the density of states of the free system, i.e. of the
Hubbard model with the coupling $U$ set to zero (a solvable system).
In this way the constant in question gets expressed in terms of the
density (in the context of the lattce models called also the degree of
filling), i.e. the ratio of the number of fermions to the number
of the lattice sites of the system and the coupling constant $U$.
In this note after recalling the steps made by Bach and Poelchau we
show that their general scheme of the derivation of the estimate can be
almost straightforwardly generalized in two directions. Firstly, the
adopted in \cite{BP} assumption about strict translational invariance of
the hopping matrix can be relaxed; secondly, a class of multiband Hubbard
models can be treated in the similar manner. The generalization 
essentially consists of modifyng the solvable part of the problem, that is
to consider the appropriate density of states and relating the mentioned
constant to the characteristics of the considered version of the
Hubbard model.


\section{Generalities}
\label{sec:generalities}

We thus consider spin one-half particles living on a finite subset
$\Lambda=[-L+1,L]^d\cap\mathbb{Z}^d$ of the hypercubic lattice. Periodic
boundary conditions in all directions will be assumed ($\Lambda$ becomes
in this way a `discrete hypertorus'). The {\em one-particle} Hilbert space
${\cal H}$ is therefore the space of spinor-valued functions:
${\cal H}=\mathbb{C}^{|\Lambda|}\otimes\mathbb{C}^2$ of (complex) dimension
$2|\Lambda|$. The Hilbert space of the simplest version of the fermionic
Hubbard model (considered in \cite{BP}) is built as a Fock space
${\cal F}({\cal H})$ over the one-particle Hilbert space ${\cal H}$, that
is as the direct sum
\begin{eqnarray}
  {\cal F}({\cal H})=\bigoplus_{N=0}^{2|\Lambda|}{\cal F}_N~\!,\nonumber
\end{eqnarray}
of $N$-particle Hilbert spaces ${\cal F}_N\equiv{\cal F}_N({\cal H})$
constructed as $N$-fold antisymmetrized tensor products of one-particle
Hilbert spaces (since the ${\cal H}$ is $2|\Lambda|$-dimensional, maximal
$N$ equals $2|\Lambda|$ and the dimension of the entire ${\cal F}({\cal H})$
is $2^{2|\Lambda|}=4^{|\Lambda|}$). The zero particle space is a one
dimensional space spanned by the no-particle ``vacuum'' vector which we
denote $|0\rangle$. In the considered case of spin 1/2 fermions the spaces
${\cal F}_N$ naturally split into direct sums of subspaces
${\cal F}_{N_+,N_-}\equiv{\cal F}_{N_+,N-N_+}$, where $N_\pm$ are numbers
of fermions with spin up and down.
As usually, with every one-particle state
$|\psi\rangle\in{\cal H}$ one can associate the creation and annihilation
operators $c^\dagger(\psi)$, $c(\psi)$ which act on ${\cal F}({\cal H})$. 
The $N$-particle Hilbert space ${\cal F}_N$ can be then spanned by
$\left(\matrix{2|\Lambda|\cr N}\right)$ vectors of the form 
\begin{eqnarray}
  \left(\prod_{k=1}^Nc^\dagger(f_{i_k})\right)|0\rangle~\!,\nonumber
\end{eqnarray}
where $(i_1,\dots,i_N)$ are all possible $N$-element subsets of the
set $2|\Lambda|$  indices $i$ labeling vectors $|f_i\rangle$ of
an orthonormal basis of ${\cal H}$. One particular such basis is
formed by vectors $|\mathbf{x},\sigma\rangle$, where $\sigma=\pm$
and $\mathbf{x}$ runs over $|\Lambda|$ sites of the lattice $\Lambda$
(the state $|\mathbf{x},\sigma\rangle$ represents the particle with
the spin projection $\sigma$ localized at the site $\mathbf{x}$).
The corresponding creation and annihilation operators
denoted $c^\dagger_{\mathbf{x},\sigma}$, $c_{\mathbf{x},\sigma}$
create and annihilate a fermion with the spin projection $\sigma$
at the site $\mathbf{x}$. Obviously, if
\begin{eqnarray}
  |f_i\rangle=\sum_{\mathbf{x}\in\Lambda}\sum_{\sigma=\pm}|\mathbf{x},\sigma\rangle~\!
  f_i^\sigma(\mathbf{x})~\!,
\end{eqnarray}
($f_i^\sigma(\mathbf{x})$ is the lattice ``wave function'' of the fermion
in the state $|f_i\rangle$), then
\begin{eqnarray}
  c^\dagger(f_i)=\sum_{\mathbf{x}\in\Lambda}\sum_{\sigma=\pm}c^\dagger_{\mathbf{x},\sigma}~\!
  f_i^\sigma(\mathbf{x})~\!,
  \phantom{aaa}
  c(f_i)=\sum_{\mathbf{x}\in\Lambda}\sum_{\sigma=\pm}c_{\mathbf{x},\sigma}~\!
  (f_i^\sigma(\mathbf{x}))^\ast~\!.\label{eqn:tradingCrAnOps}
\end{eqnarray}

In the language of second quantization the Hamiltonian of the
simplest version of the Hubbard model, written in terms
of the operators $c^\dagger_{\mathbf{x},\sigma}$, $c_{\mathbf{x},\sigma}$,
takes the form
\begin{eqnarray}
\hat H=\hat T+\hat V~\!,\label{HamHuM}
\end{eqnarray}
with
\begin{eqnarray}
\hat T=-\sum_{\langle\mathbf{x},\mathbf{y}\rangle}\sum_{\sigma=\pm} t_{\mathbf{x},\mathbf{y}}~\!
c^\dagger_{\mathbf{x}\sigma} c_{\mathbf{y}\sigma}~\!,\label{Tkin}
\end{eqnarray}
and
\begin{eqnarray}
\hat V=U\sum_{\mathbf{x}} n_{\mathbf{x}+}n_{\mathbf{x}-}~\!.\label{Vint}
\end{eqnarray}
Here $t_{\mathbf{x}\mathbf{y}}$ is the hopping matrix, $U$ is the Coulomb coupling
constant (we consider a repulsive interaction with $U>0$) and
$n_{\mathbf{x}\sigma}=c^\dagger_{\mathbf{x}\sigma}c_{\mathbf{x}\sigma}$ are the
operators of the numbers of particles with the spin projection $\sigma$ 
at the lattice site $\mathbf{x}$. The symbol
$\sum_{\langle\mathbf{x}\mathbf{y}\rangle}$ means summation over all possible
pairs of lattice sites. In this Section, following Bach and Poelchau,
we assume that the hoping matrix $t_{\mathbf{x}\mathbf{y}}$ is translationally
invariant that is, that
$t_{\mathbf{x},\mathbf{y}}=t_{\mathbf{x}-\mathbf{y},\mathbf{0}}\equiv t(\mathbf{x}-\mathbf{y})$.
(In the following this assumption will be modified by allowing for periods 
greater than the unit one).

It will be convenient to introduce also the operators
\begin{eqnarray}
  \hat{\mathbf{N}}=\sum_{\mathbf{x}}\sum_{\sigma=\pm} n_{\mathbf{x},\sigma}~\!,
  \phantom{aaa}{\rm and}
  \phantom{aaa}\hat{\mathbf{n}}={\hat{\mathbf{N}}\over|\Lambda|}~\!,
\label{OpCalkLiCzast}
\end{eqnarray}
of the {\em total} number of particles and of the total density,
respectively
\vskip0.2cm

We first define the true {\it ground state energy} $E_{gs}(n)$ of the
$N$-fermion system as the infimum of the expectation values of the
Hamiltonian (\ref{HamHuM}) in states belonging to ${\cal F}_N$
that is, in states of the system of density $n=N/|\Lambda|$:
\begin{eqnarray}
E_{gs}(n)={\rm inf}\{\langle\Psi|\hat H|\Psi\rangle:\phantom{aa}
|\Psi\rangle\in{\cal F}({\cal H}),\phantom{aa}\langle\Psi|\Psi\rangle=1,
\phantom{aa}\hat{\mathbf{n}}|\Psi\rangle=n|\Psi\rangle\}~\!.\label{EgsAsInf}
\end{eqnarray}
As the Hilbert space of the considered system is finite-dimensional,
$E_{gs}(n)$ given by (\ref{EgsAsInf}) is the same as the lowest eigenvalue
in ${\cal F}_N$ of the Hamiltonian $\hat H$, that is its eigenvalue on a
(normalized) {\em ground-state vector} $|\Psi_{gs}\rangle\in{\cal F}_N({\cal H})$
such that $\hat{\mathbf{n}}|\Psi_{gs}\rangle=n|\Psi_{gs}\rangle$ and
$\hat H|\Psi_{gs}\rangle=E_{gs}|\Psi_{gs}\rangle$. The formula
(\ref{EgsAsInf}) is, however, more convenient in further considerations
than the eigenvalue equation.
 
The corresponding {\em Hartree-Fock (H-F) ground state energy} $E_{hf}(n)$
is defined as
\begin{eqnarray}
  E_{hf}(n)={\rm inf}\{\langle\Phi|\hat H|\Phi\rangle~\!,\quad
  |\Phi\rangle\in{\cal S}{\cal D}_N\}~\!.\label{HFenergy}
\end{eqnarray}
${\cal S}{\cal D}_N\subset{\cal F}_N({\cal H})$
denotes the set of $N$-particle vectors constructed as
the {\em Slater determinants} that is, of vectors of the form
\begin{eqnarray}
{\cal S}{\cal D}_N=\left\{|\Phi\rangle=\prod_{j=1}^N c^\dagger(f_j)|0\rangle:\quad
  |f_1\rangle,\dots,|f_N\rangle\in{\cal H}:\quad \langle f_i|f_j\rangle=\delta_{ij}
 \right\}~\!.\label{SlaterDet}
\end{eqnarray}
A vector $|\Phi_{hf}\rangle\in{\cal S}{\cal D}_N$ such that
$E_{hf}=\langle\Phi_{hf}|\hat H|\Phi_{hf}\rangle$
will be called the {\em Hartree-Fock ground state} of the Hamiltonian $\hat H$.

Both quantities, $E_{hf}(n)$ and $E_{gs}(n)$, are thus obtained by minimizing
the expectation values of the same Hamiltonian $\hat H$ (\ref{HamHuM}) but
each one over a different set of vectors (belonging to ${\cal F}_N$):
$E_{gs}(n)$ is the minimum over
the entire $N$-particle subspace ${\cal F}_N$, that
is over all possible vectors $|\Psi\rangle$ which in general have the form
\begin{eqnarray}
  |\Psi\rangle=\sum_{\mathbf{x}_1,\dots,\mathbf{x}_N\in\Lambda}
  \sum_{\sigma_1,\dots,\sigma_N}\psi_{\sigma_1,\dots,\sigma_N}(\mathbf{x}_1,\dots,\mathbf{x}_N)
  ~\!c^\dagger_{\mathbf{x}_1\sigma_1}\dots c^\dagger_{\mathbf{x}_N\sigma_N}|0\rangle~\!,
  \nonumber
  \end{eqnarray}
while $E_{hf}(n)$ is the minimum over only the set of $N$-particle vectors
having the factorizable form, that is (cf. the formula
(\ref{eqn:tradingCrAnOps})) such that\footnote{The antisymmetrization
  which is implied by the name ``Slater determinant'' is, of course, ensured
  by the properties of the operators $c^\dagger_{\mathbf{x},\sigma}$.}
\begin{eqnarray}
  \psi_{\sigma_1,\dots,\sigma_N}(\mathbf{x}_1,\dots,\mathbf{x}_N)
  =f_1^{\sigma_1}(\mathbf{x}_1)\cdot\dots\cdot f_N^{\sigma_N}(\mathbf{x}_N)~\!.
  \nonumber
\end{eqnarray}
The standard Ritz variational principle implies therefore the inequality
\begin{eqnarray}
E_{gs}(n)\leq E_{hf}(n)\quad {\rm i.e.}\quad
\Delta E(n)\equiv E_{gs}(n)-E_{hf}(n)\leq 0~\!.\label{eqn:DeltaEdef}
\end{eqnarray}
In other words, $\Del E(n)$ is by construction bounded from {\em from above}.
\vskip0.1cm

In \cite{BP} a {\em lower bound} on $\Del E(n)$ has been obtained in 
the form of the following

\noindent{\bf Theorem} \cite{BP}. Let $t$ be a translationally invariant
hopping matrix with a finite support, such that in the thermodynamic (TD)
limit its Fourier transform $\hat t\equiv\epsilon$ (that is, the
{\em dispersion function}), $\epsilon:[-\pi,\pi]^d\rightarrow\mathbb{R}$,
is a Morse function (i.e. a function
the critical points of which are non-degenerate), and let $\Lambda$ be
a sufficiently large lattice such that $\rm{supp}(t)\subset\La$. Finally,
let $n=N/\Lambda$ be the density. Then:\\
$\bullet$ if the dimension of the lattice $\Lambda$ is $d=2$, 
\begin{eqnarray}
  0\geq \Del E(n)\geq -{\rm const.}\left(n^{2/3}U^{4/3}(1+|\ln U|)
  + n^{1/2}U |\La|^{-1/4}(1+\ln(|\Lambda|^{-1/2}))\right),
\end{eqnarray}
$\bullet$ if the dimension of the lattice $\Lambda$ is $d\geq 3$,
\begin{eqnarray}
  0\geq\Del E(n)\geq -{\rm const.}\left(n^{2/3}U^{4/3}
  + n^{1/2}U|\Lambda|^{-1/2d}\right).
\end{eqnarray}
\hfill $\blacksquare$
 
This theorem has been proved by using the {\em correlation inequalities}.
We present main points of this proof below. After that, we present our
extensions of this result.

\section{Sketch of the proof}

To present the sketch of the proof some additional definitions must be
given. We first introduce the family of projection operators
\begin{eqnarray}
\hat X_{\mathbf{x}}=\sum_\sigma |\mathbf{x},\sigma\rangle
\langle\mathbf{x},\sigma|~\!,\label{Xczylin}
\end{eqnarray}
acting on ${\cal H}$ and such that ($\hat1_1$ is the unit operator
acting on ${\cal H}$)
\begin{eqnarray}
\sum_{\mathbf{x}} \hat X_{\mathbf{x}} = \hat1_1~\!.\label{SumXx}
\end{eqnarray}
The interaction operator (\ref{Vint}) restricted to the
  two-particle space ${\cal F}_2$ and extended to act on the
  entire ${\cal H}\otimes{\cal H}$ (denoted therefore $\hat V_2$)
can be expressed in terms of $\hat X_{\mathbf{x}}$ as:
\begin{eqnarray}
\hat V_2=U\sum_{\mathbf{x}}\hat X_{\mathbf{x}}\otimes\hat X_{\mathbf{x}}~\!, 
\label{V2term}
\end{eqnarray}
Similarly, the kinetic energy operator (\ref{Tkin}) restricted
to the one-particle space ${\cal F}_1\equiv{\cal H}$ (and denoted
therefore $\hat T_1$) can be written as
\begin{eqnarray}
  \hat T_1 =\sum_{\langle\mathbf{x},\mathbf{y}\rangle}\sum_\sigma
  t_{\mathbf{x},\mathbf{y}}|\mathbf{x},\sigma\rangle\langle\mathbf{y},\sigma|~\!.
\end{eqnarray}

Needed will be also the {\em density operators} which can be defined as
follows. Let the set of vectors $|\phi_1\rangle,\dots,|\phi_{2|\La|}\rangle$
form an orthonormal basis of ${\cal H}$ and let $|\Psi\rangle$ be some
normalized vector of ${\cal F}({\cal H})$. The associated operators:
the {\bf one-particle density operator} (1-pdo) $\hat\gamma_\Psi$
acting on ${\cal H}$,
and the {\bf two-particle density operator} (2-pdo)  $\hat\Gamma_\Psi$ 
acting on ${\cal H}\otimes{\cal H}$ are then defined by the formulae
\begin{eqnarray}
  \hat\gamma_\Psi=\sum_{k,l}|\phi_l\rangle~\!
  \langle\Psi|c^\dagger(\phi_k) c(\phi_l)|\Psi\rangle~\!
  \langle\phi_k|~\!,\label{eqn:1pdo}\phantom{aaaaaaaaaa}\\
  \hat\Gamma_\Psi = \sum_{k,l,m,n}|\phi_m\rangle\otimes|\phi_n\rangle~\!
  \langle\Psi|c^\dg(\phi_k)c^\dg(\phi_l) c(\phi_m)c(\phi_n)|\Psi\rangle~\!
  \langle\phi_k|\otimes\langle\phi_l|~\!.\label{eqn:2pdo}
\end{eqnarray}
(Higher-particle densities can be defined analogously, but will not
be needed in our considerations). The operators $\hat\gamma_\Psi$ and
$\hat\Gamma_\Psi$ have the following, easy to check, properties:
\begin{itemize}
\item They are independent of the choice of the orthonormal basis
  $|\phi_i\rangle$ of ${\cal H}$
\item They are both positive Hermitian operators such that: 
\begin{eqnarray}
0\leq\hat\gamma_\Psi=\hat\gamma^\dagger_\Psi\leq\hat1_1~\!,\phantom{aaa}
{\rm Tr}_1(\hat\gamma_\Psi)=\langle\Psi|\hat N|\Psi\rangle=N~\!,\nonumber\\ 
0\leq\hat\Gamma_\Psi=\hat\Gamma^\dagger_\Psi\leq\hat1_2~\!
{\rm Tr}_2(\hat\Gamma_\Psi)~\!,
\phantom{aaaaaaaaa}
\end{eqnarray}
(Tr$_1$ and Tr$_2$ denote traces of operators acting on ${\cal H}$ and
  ${\cal H}\otimes{\cal H}$, respectively and $\hat1_1$ and $\hat1_2$
  are the unit operators on these subspaces).
\item If $|\Phi\rangle\in{\cal S}{\cal D}_N$, i.e. if the state $|\Phi\rangle$
is constructed as a Slater determinant, then $\hat\gamma_\Phi=\hat\gamma^2_\Phi$
and, moreover, $\hat\gamma_\Phi$ is an orthogonal projection
onto the $N$ dimensional subspace of ${\cal H}$ spanned by the vectors used to
construct $|\Phi\rangle$.
\item The 2-pdo associated with $|\Phi\rangle\in{\cal S}{\cal D}_N$ takes
  the form
$\hat\Gamma_\Phi=\hat A_2(\hat\gamma_\Phi\otimes\hat\gamma_\Phi)$,
where $\hat A_2$ is the {\em antisymmetrization operator} acting on
${\cal H}\otimes{\cal H}$: $\hat A_2|f\rangle\otimes|g\rangle
=|f\rangle\otimes|g\rangle-|g\rangle\otimes|f\rangle$ (in the notation
of \cite{BP} $\hat A_2=\hat1_2-\hat{Ex}$, where $\hat{Ex}$ is the
exchange operator).
\end{itemize}

Let now $\hat\gamma_{gs}$ and $\hat\Gamma_{gs}$ be the 1- and 2-pdo's,
respectively,
corresponding to the true ground state $|\Psi_{gs}\rangle$ of the Hubbard model
Hamiltonian (\ref{HamHuM}). The ground state energy $E_{gs}$ can be then
written as
\begin{eqnarray}
  E_{gs}(n) = {\rm Tr}_1(\hat T_1\hat\gamma_{gs})
  +\frac{1}{2}{\rm Tr}_2(\hat V_2\hat\Gamma_{gs})~\!.
\end{eqnarray}
where $\hat T_1$ and $\hat V_2$ have been defined above.
In the analogous manner the H-F ground state energy can be written as
\begin{eqnarray}
E_{hf}(n) = {\rm inf}\left\{{\cal E}_{HF}[\ga_\Phi],
  \phantom{aa}|\Phi\rangle\in{\cal S}{\cal D}_N\right\},\label{HFenergy2form}
\end{eqnarray}
where
\begin{eqnarray}
  {\cal E}_{HF}[\ga_\Phi]\equiv{\rm Tr}_1(\hat T_1\hat\gamma_{\Phi})
    +{1\over2}{\rm Tr}_2[\hat V_2
    \hat A_2(\hat\gamma_\Phi\otimes\hat\gamma_\Phi)],\nonumber
\end{eqnarray}
is  the {\em the Hartree-Fock energy functional}.
\vskip0.1cm

The lower bound on $\Delta E(n)$ defined in (\ref{eqn:DeltaEdef})
is now obtained as follows.
Let $\hat\gamma_0$ be the 1-pdo associated with the ground state
$|\Phi_0\rangle$ of the free Hamiltonian, i.e. of $\hat T_1$ (\ref{Tkin}).
Obviously $|\Phi_0\rangle$ has the form of a Slater determinant.
One expects that $\hat\gamma_{gs}(U)\rightarrow\hat\gamma_0$ for
$U\rightarrow0$. The rigorous proof of this fact in the form
of a lower bound 
on $\Delta E(n)$ has been given in \cite{BP}. The bound, being
a function of the coupling $U$, asseses
the rate of this convergence.
The proof consists of the following points:
i) write an estimate of $\Delta E(n)$ using
$\hat\Gamma_{gs}$ (Lemma 1); ii) using the appropriate correlation
inequality, go over from $\hat\Gamma_{gs}$ to $\hat\gamma_{gs}$
(reduction of the two-body problem to a one-body one) (Lemma 2);
iii) put a bound on $\hat\gamma_{gs}$ to obtain the final estimate of
$\Delta E(n)$ in terms of $U$, $n$ and $|\La|$ only (Lemmas 3 and 4).
More explicitly we have \cite{BP}:\\
\noindent{\bf Lemma 1.} 
\begin{eqnarray}
\Delta E(n) \geq \frac{U}{2}\sum_{\mathbf{x}\in\Lambda}
{\rm Tr}_2\!\left[(\hat X_{\mathbf{x}}\otimes\hat X_{\mathbf{x}})\left(
\hat\Gamma_{gs}-\hat A_2(\hat\gamma_0\otimes\hat\gamma_0)\right)\right].
\label{IneqL1}
\end{eqnarray}
\hfill $\blacksquare$

\noindent{\bf Lemma 2.} (\cite{GrafSolovej}; $\hat X$ stands for a 
bounded Hermitian projection operator acting on ${\cal H}$)
\begin{eqnarray}
  \frac{1}{2}{\rm Tr}_2[(\hat X\otimes\hat X)\hat\Gamma_{gs}]
  -{\rm Tr}_1[\hat X\hat\gamma_0]{\rm Tr}_1[\hat X\hat\gamma_{gs}]
  +\frac{1}{2}\left({\rm Tr}_1[\hat X\hat\gamma_0]\right)^2
  +\frac{1}{2}{\rm Tr}_1[\hat X\hat\gamma_0\hat X\hat\gamma_0]\geq
  \phantom{aaaaa}\label{Lemma2}\\
  -~\!{\rm const.}({\rm T}_1[\hat X\hat\gamma_0]
  +{\rm Tr}_1[\hat X\hat\gamma_{gs}]+{\rm Tr}_1[\hat X\hat\gamma_0])\cdot
  {\rm min}\!\left(1, \sqrt{{\rm Tr}_1[\hat X(\hat1_1-\hat\gamma_0)
      \hat\gamma_{gs}(\hat1_1-\hat\gamma_0)]}\right).\nonumber
\end{eqnarray}
\hfill $\blacksquare$

\noindent{\bf Lemma 3.} 
\begin{eqnarray}
\frac{\Delta E(n)}{|\Lambda|}\geq
-{\rm const.}~\!U~\!\sqrt{n}~\!\sqrt{\frac{{\rm Tr}_1[(\hat1_1-\hat\gamma_{gs})
      \hat\gamma_0]}{|\Lambda|}}\equiv
-{\rm const.}~\!U~\!\sqrt{n}~\!A~\!.\label{IneqL3}
\end{eqnarray}
\hfill $\blacksquare$

The final element of the proof is to find an estimate of the constant $A$.
It can be expressed
in terms of the spectrum of the noninteracting ($U=0$) system, i.e.
on the eigenvalues of the hopping matrix $t_{\mathbf{x},\mathbf{y}}$ and
the corresponding {\rm density of states} (DOS). Since the hopping matrix has
been assumed to be translationally invariant, it can be diagonalized with the
help of the (discrete) Fourier transform. This brings the
$\hat T$ operator (and, hence, the free Hamiltonian) into the diagonal form:
\begin{eqnarray}
  \hat T=\sum_{\mathbf{k}\in{\cal B}{\cal Z}}\sum_\sigma\epsilon(\mathbf{k})~\!
  c^\dagger_{\mathbf{k},\sigma}c_{\mathbf{k},\sigma}~\!,
\label{ek}
\end{eqnarray}
where ${\cal B}{\cal Z}=[-\pi,\pi)^d\cap(\pi/L)\mathbb{Z}^d$
is the first {\em Brillouin zone}, the function $\epsilon(\mathbf{k})$
(the `free-particle spectrum', or the `dispersion relation') is the
discrete Fourier transform of the hopping matrix:
\begin{eqnarray}
  \epsilon(\mathbf{k})=\hat t(\mathbf{k})=\sum_{\mathbf{x}\in\Lambda}
  t(\mathbf{x})~\!e^{i\mathbf{k}\cdot\mathbf{x}}~\!,\label{ek2}
\end{eqnarray}
and the operators $c^\dagger_{\mathbf{k},\sigma}$ and $c_{\mathbf{k},\sigma}$ are given by
\begin{eqnarray}
  c^\dagger_{\mathbf{k},\sigma}={1\over\sqrt{|\Lambda|}}\sum_{\mathbf{x}\in\Lambda}
  e^{i\mathbf{k}\cdot\mathbf{x}}~\!c^\dagger_{\mathbf{x},\sigma}~\!, 
\phantom{aaa}
  c_{\mathbf{k},\sigma}=\frac{1}{\sqrt{|\Lambda|}}\sum_{\mathbf{x}\in\Lambda}
  e^{-i\mathbf{k}\cdot\mathbf{x}}~\!c_{\mathbf{x},\sigma}~\!. 
\end{eqnarray}
Of course, the vectors
$|\mathbf{k},\sigma\rangle=c^\dagger_{\mathbf{k},\sigma}|0\rangle$
form an orthonormal basis of ${\cal H}$ and the ground state
$|\Psi_0\rangle$ (belonging to the ${\cal F}_{N_+,N_-}$ subspace)
of the $U=0$ Hubbard model can be, assuming that
$\epsilon(\mathbf{k}_j)\leq\epsilon(\mathbf{k}_i)$ for $j<i$,
written as ($N_++N_-=N$)
\begin{eqnarray}
  |\Psi_0\rangle=\left(\prod_{j_1=1}^{N_+}
  c^\dagger_{\mathbf{k}_{j_1},+}\right)\! \left(\prod_{j_2=1}^{N_-}
  c^\dagger_{\mathbf{k}_{j_2},-}\right)\!|0\rangle~\!.
\end{eqnarray}
This allows to introduce the {\em Fermi energy} $\epsilon_{\rm F}$
as the highest energy of the occupied one-particle states, that is
$\epsilon_{\rm F}=\epsilon(\mathbf{k}_{j_{\rm max}})$, where
$j_{\rm max}={\rm max}(N_+,N_-)$.

The density of states (of the non-interacting system, $U=0$), 
is defined by the formula
\begin{eqnarray}
  \rho_d(E) =\int_{[-\pi,\pi]^d}\df^d\mathbf{k}~\!
  \delta(\epsilon(\mathbf{k})-E)\equiv{\df N_d(E)\over\df E}~\!,\label{eqn:DOS}
\end{eqnarray}
where
\begin{eqnarray}
N_d(E) =\int_{[-\pi,\pi]^d}\df^d\mathbf{k}~\!\chi(\eps(\mathbf{k})\leq E)~\!.
\end{eqnarray}
The necessary estimate of the constant $A$ appearing in the inequalty
(\ref{IneqL3}) is now provided by\\
\noindent {\bf Lemma 4.} (the `bootstrap' estimate).\\
\noindent There exist constants $c_1$, $c_2$, $c_3$ such that for every
$\varepsilon>0$ the constant $A$ satisfies the bound ($2L=|\Lambda|^{1/d}$
is the width of the lattice):
\begin{eqnarray}
A^2\leq{c_1\over\varepsilon}~\!U A\sqrt{n}
+ c_2\!\int^{c_3/L}_{-\veps-c_3/L}\!\df\lambda~\!\rho_d(\epsilon_{\rm F}+\lambda)
\equiv{c_1\over\varepsilon}~\!U A\sqrt{n}
+ c_2~\!I(\varepsilon,L;n) ~\!.
\label{BootstrapEst}
\end{eqnarray}
\hfill $\blacksquare$

Solving the inequality (\ref{BootstrapEst}) one obtains the upper bound
on $A$
\begin{eqnarray}
  A\leq{c_1\over2\varepsilon}~\!U\sqrt{n}+\sqrt{\left(
{c_1\over2\varepsilon}~\!U\right)^2n+c_2~\!I(\varepsilon,L;n)}~\!.
\label{BootstrapSolved}
\end{eqnarray}
Combined with the inequality (\ref{IneqL3}), this bound provides the
general form (obtained in \cite{BP}) of the the lower bound on the
difference (\ref{eqn:DeltaEdef}) of the true ground state energy and the
H-F approximation to it. The precise form of this bound, i.e. its final
dependence on $n$ and $U$ (and on the lattice dimension $d$), depends on
the estimate of the integral $I(\varepsilon,L;n)$. The latter depends
on the general form of density of states $\rho_d(E)$ (which is a property
of the type of the lattice and the number of dimensions $d$) and also on
the filling of the lattice. Estimating $I(\varepsilon,L;n)$ is
straightforward in those cases in which $\rho_d(E)$ is a bounded function;
in some cases it has, however, (integrable) singularities which introduce
some complications in estimating $I(\varepsilon,L;n)$. These will be
discussed in the next Section.

\section{Examples of estimates}

In this Section we show how the machinery developed in \cite{BP} and
presented above works in practice. In Sections \ref{susec:d3hypercubic}
and \ref{subsec:Square} we repeat (for the illustration) the derivation
of the estimates obtained in \cite{BP}, and in the subsequent two sections
we present our extension of this approach to the versions of the Hubbard
models not considered in \cite{BP}: to the single band model on the
body centered cubic (bcc) lattice and to the flat band systems.

\subsection{A prototype calculation: the (hyper)cubic lattice in $d\geq3$}
\label{susec:d3hypercubic}

Let the elements of the hopping matrix be non-zero and equal $t$ only for
the nearest neighbour sites of the $d$-dimensional simple (hyper)cubic
lattice (sc). The dispersion function has then the form
\begin{eqnarray}
\epsilon_{sc}(\mathbf{k}) =-2t~\!(\cos k_1+\dots+\cos k_d)~\!.
\label{scDispersion}
\end{eqnarray}
It is therefore a Morse function (i.e. it has only non-degenerate
critical points). For $d\geq3$ the corresponding DOS function
(\ref{eqn:DOS}), which is given by a non-elementary integral \cite{HUMH},
is bounded and the integral $I(\varepsilon,L;n)$ defined by (\ref{BootstrapEst})
can be easily estimated ($c_4={\rm max}\{4c_3,1\}$):
\begin{eqnarray}
I(\varepsilon,L;n)\leq\left({2c_3\over L}+\varepsilon\right){\rm sup}(\rho_d)
\equiv c_4\left(|\Lambda|^{1/d}+\varepsilon\right){\rm sup}(\rho_d)~\!.\nonumber
\end{eqnarray}
The resulting bound (\ref{BootstrapSolved}) 
\begin{eqnarray}
A\leq{c_1\over2\varepsilon}~\!U\sqrt{n}+
\sqrt{{c_1^2 U^2n\over4\varepsilon^2}+c_4\varepsilon+c_4|\Lambda|^{1/d}}~\!
\nonumber
\end{eqnarray}
is valid for any $\varepsilon>0$ and can be, therefore, brought into a
reasonably optimal form (using the inequality $\sqrt{a+b}<\sqrt{a}+\sqrt{b}~\!$)
by simply setting $\varepsilon={\rm const.}~\!n^{1/3} U^{2/3}$:
\begin{eqnarray}
A\leq {\rm const.}\left(n^{1/6} U^{1/3}+{\rm const.}^\prime|\Lambda|^{1/d}
\right).\nonumber
\end{eqnarray}
Combining this with the relation (\ref{IneqL3})
we obtain the lower bound
\begin{eqnarray}
{\Delta E(n)\over|\Lambda|}
\geq-{\rm const.}\left(n^{2/3} U^{4/3}+{\rm const.}^\prime
n^{1/2} U |\Lambda|^{1/d}\right).\label{eqn:FirstBound}
\end{eqnarray}

\subsection{The square lattice in $d=2$}
\label{subsec:Square}

In the case of the two-dimensional square lattice with a nonzero hopping
amplitude for nearest neighbour sites only the density of states is given
by a complete elliptic integral of the first kind (\cite{HUMH}:
\begin{eqnarray}
  \rho(\varepsilon)={1\over2|t|\pi^2}~\!K\!\!\left(1-(\epsilon/4t)^2\right),
  \nonumber
\end{eqnarray}
(the plot is given in Figure 1a of this reference). From the well known
properties of the elliptic integral $K(k)$ which logarithmically diverges
at $k=1$ it follows that $\rho(\epsilon)$ has such a singularity at
$\epsilon=0$. Depending on the value of the Fermi energy
$\epsilon_{\rm F}$, which is determined by the filling, the integration
domain in $I(\veps,L;n)$ may or may not contain this singularity. In the
latter case, the density of states is (within the integration domain)
bounded and essentially the same estimate of $I(\varepsilon,L;n)$ as
in the case of the hypercubic lattice in $d\geq 3$. The condition
  on density $n$ which distinguishes this case from the other one
is rather involved. We therefore do not show details
of this analysis and concentrate on the opposite case of sufficiently
high filling in which the presence of the singularity affects the
estimate of $I(\epsilon,L;n)$. 

It this case, using the asymptotic form of the elliptic relevant integral,
one can write
\begin{eqnarray}
  I(\varepsilon,L;n)\leq C\left[\left(\frac{c}{L}+\varepsilon\right)\!
  \ln\!\left(\frac{c}{L}+\varepsilon\right)+\varepsilon
  + \frac{c}{L}\ln\frac{c}{L} + 2\frac{c}{L}\right].
\end{eqnarray}
Setting here $\varepsilon={\rm Const.}~\!n^{1/3}U^{2/3}$ (which again is
a simple but reasonable choice) we obtain as the upper bound 
\begin{eqnarray}
  A\leq C^{\prime\prime}\!\left\{n^{1/6}U^{1/3}\matrix{\phantom{a}\cr
    \phantom{a}}\right.
  \phantom{aaaaaaaaasssssssssssssaaaaaaaaaaaaaaaaaaaaaaaaaa}~\nonumber\\
  \left.+\sqrt{n^{1/3}U^{2/3}
    +\left(\frac{1}{L}+n^{1/3} U^{2/3}\!\right)\!\ln\!\left(\frac{1}{L}
    +n^{1/3} U^{2/3}\!\right)+\frac{1}{L}\ln\frac{1}{L} +\frac{2}{L}}~\!
\right\}.\label{eqn:ComplictedEstimate}
\end{eqnarray}
In conjunction with the inequality (\ref{IneqL3}) this
suffices to conclude that for small $U$ and large $|\Lambda|$ the H-F
approximation becomes exact. We note also that
the bound on $A$ given in \cite{BP} for this particular case,
although more transparent, is slightly less stringent than the one
given above. Nevertheless, it too leads to the same overall conclusion.
This bound follows from the one given above if one takes
into account that the constraint $0\leq n\leq 2$ implies
boundedness of $n\ln n$; using the inequality
$\sqrt{a+b}\leq \sqrt{a}+\sqrt{b}$ after a bit lengthy but
elementary algebra one arrives at
\begin{eqnarray}
A\leq C\left\{n^{1/6}U^{1/3}(\ln|U|+1) + |\Lambda|^{-1/4}
 |\ln(|\Lambda|^{-1/2}|+1)\right\},
\end{eqnarray}
which leads to
\begin{eqnarray}
\frac{\Delta E(n)}{|\Lambda|}\geq
-C\left[n^{2/3}U^{4/3}(\ln|U|+1) + n^{1/2}U|\Lambda|^{-1/4}
|\ln(|\La|^{-1/2}|+1)\right].\label{eqn:SeconBound}
\end{eqnarray}
We remark that the above estimate is universally aplicable at
all densities (fillings) in view of the fact that it is weaker
than the one obained for high densities.

\subsection{Body-centered cubic lattice in $d=3$}

The body-centered cubic lattice (bcc) is an example of a lattice on which
the period of the hopping matrix $t_{\mathbf{x},\mathbf{y}}$ is greater than 1.
Application to this case of the machinery developed in \cite{BP} requires
again its slight generalization. Here we consider only the simplest case
that is, we assume that the hopping constants (nonzero entries of the matrix
$t_{\mathbf{x},\mathbf{y}}$) are all equal $t$. The dispersion relation which
originates from the periodic structure of such a bcc lattice is \cite{HUMH}
\begin{eqnarray}
\epsilon_{bcc}(\mathbf{k}) = -8t\cos k_1 \cos k_2 \cos k_3~\!.
\label{bccDispersion}
\end{eqnarray}
The corresponding DOS function $\rho_{bcc}(\epsilon)$ cannot be expressed
in terms of elementary functions. It is nevertheless possible to extract
those its properties which are important for estimating the integral
$I(\varepsilon,L;n)$. It turns out that the DOS function in question does
have a {\em singularity}: near the zero energy
$\rho_{bcc}(\epsilon)\sim(4\pi)^{-3}\ln^2\epsilon$ \cite{HUMH}.
The singularity originates from the neighbourhood of the point
$(\pi/2,~\!\pi/2,~\!\pi/2)$ at which all the cosine functions
in (\ref{bccDispersion}) simultaneously vanish. In estimating the
integral $I(\varepsilon,L;n)$, assuming that the singularity
of the DOS function lies within the integration domain, we follow,
therefore, the steps taken in \cite{BP} in deriving the estimate for
the $d=2$ square lattice (in which case the DOS function has also a
logarithmic singularity, $\sim\ln\varepsilon$).
The estimate of the integral $I(\varepsilon,L;n)$  obtained using
the asymptotic form of $\rho_{bcc}$ reads ($c$ is some constant):
\begin{eqnarray}
I(\varepsilon, L;n) \leq 
{c\over L}\!\left\{4+\ln^2\!{c\over L} + 2\left|\ln{c\over L}\right|\right\}
+ 2\varepsilon
\phantom{aaaaaaaaaaa}\nonumber\\
+\left({c\over L}+\varepsilon\right)\!\left\{
\ln^2\!\left(\varepsilon+{c\over L}\right)+2\left|
\ln\!\left(\varepsilon+{c\over L}\right)\right|\right\}.
\end{eqnarray}
Seting again $\varepsilon = {\rm Const.}^\prime n^{1/3} U^{2/3}$ we get a
reasonably
stringent, but unfortunately having a rather complicated form, bound
on $A$ as a function of $L$, $n$ and $U$ (it is somewhat analogous
to (\ref{eqn:ComplictedEstimate})). Instead of reproducing it
here we will content ourselves with a less stringent but more transparent
bound which, combined with the inequality (\ref{IneqL3}),
leads to the lower bound
\begin{eqnarray}
\frac{\Delta E(n)}{|\Lambda|}\geq
-C\left[ n^{2/3}U^{4/3}(1+\ln^2|U|+\ln|U|)
  + n^{1/2}U\times{\cal O}(|\Lambda|^{-1/6})\right].\label{eqn:ThirdBound}
\end{eqnarray}

\subsection{Flat-band systems}

We now turn to systems the dispersion functions $\epsilon(\mathbf{k})$
of which are completely flat, i.e. independent of $\mathbf{k}$.
(In general, the dispersion relations of such systems consist of several
subbands; some of them may be non-constant, but at least one is).
Such lattices play a crucial role in the emergence of the
{\em flat-band ferromagnetism} \cite{TasakiFBFM}. Perhaps, the most famous
one is the Kagom\'e lattice. Another one is the simple one-dimensional
lattice, called the {\em sawtooth} chain. A completely flat subband with
a macroscopic degeneracy of the ground state (the number of 
states having the lowest possible value of energy is proportional to
$|\Lambda|$)
implies that the corresponding DOS function has the delta function form
with a spectral weight $w$ ($w=1/3$ for the Kagom\'e lattice, $w=1/2$
for the sawtooth one). 

The fundamental inequality (\ref{BootstrapEst}) takes in this case the form
\begin{eqnarray}
A^2\leq{c_1 U \sqrt{n}\over\varepsilon}~\!A + w|\Lambda|~\!,\nonumber
\end{eqnarray}
and implies the upper bound
\begin{eqnarray}
  A\leq{c_1 U \sqrt{n}\over2\varepsilon}
  + \sqrt{\left({c_1 U \sqrt{n}\over2\varepsilon} \right)^2
    +4w|\Lambda|}~\!.\nonumber
\end{eqnarray}
Optimizing again with respect to $\varepsilon$ by setting $\varepsilon=\infty$
 we get
\begin{eqnarray}
A \leq 2\sqrt{w|\Lambda|}~\!,\nonumber
\end{eqnarray}
which, in conjunction with (\ref{IneqL3}), leads to the estimate
\begin{eqnarray}
{\Del E(n)\over|\Lambda|} \geq - {\rm Const.}~\!U \sqrt{n|\Lambda|}~\!.
\end{eqnarray}
Although correct, this result is clearly useless. The lesson that,
nevertheless, can
be drawn from this example is: {\em the more singular is the DOS
function, the worse is the estimate of} $\Delta E$.

\section{Multiband models}

In the common opinion the single-band version of the Hubbard model
considered in the preceding sections is too simple to be capable of
capturing all relevant features of realistic systems exhibiting
ferromagnetism. One is therefore naturally led to consider various
extensions of the original Hubbard model which allow to introduce a
variety of interaction terms in order to model both charge and magnetic
- e.g. the Hund one - interactions \cite{Oles,MetallicFM}. In this Section
we point out that the technique of Bach and Poelchau \cite{BP} can be
immediately extended to give estimates of the ground state energies of at
least two simplest such extensions: the $M$-band
model with the coulombic interactions only and the $SU(M)$ symmetric
extension of the single-band model which correspond to
spin $M$ fermions the interactions of which are spin independent
($SU(M)$ is then an ``accidental'' extension of the usual $SU(2)$
symmetry associated with the usual rotational invariance). Up to now
we were not able to apply the same approach to models with the
Hund-type interactions. 

\subsection{The $M$-band Hubbard Model}

Allowing for $M$-bands enlarges the dimension of the single particle
Hilbert space ${\cal H}$ to $2M|\Lambda|$ so that now the dimension
of the Fock space ${\cal F}({\cal H})$ constructed in the same way
as in Section \ref{sec:generalities} is $2^{2M|\Lambda|}$. The natural basis
of ${\cal H}$ is in this case formed by the state-vectors
$|\mathbf{x},a,\sigma\rangle$ where $a=1,\dots,M$ is the band label,
and  $\sigma=\pm$ (as previously we consider spin 1/2 fermions)
is the spin label. In terms of the creation and annihilation
operators associated with this basis of ${\cal H}$ the Hamiltonian
of the considered $M$-band version of the Hubbard model reads
\begin{eqnarray}
H=-\sum_{\langle\mathbf{x},\mathbf{x}\rangle}\sum_{a=1}^M\sum_{\sigma=\pm}
t_{\mathbf{x},\mathbf{x}}~\!c^\dagger_{\mathbf{x},a,\sigma} c_{\mathbf{y},a,\si}
+ U\sum_{\mathbf{x}}\sum_{a=1}^M n_{\mathbf{x},a,+}n_{\mathbf{x},a,-}\nonumber\\
+ U^\prime\sum_{\mathbf{x}}\sum_{a\ne a^\prime=1}^M\sum_{\sigma,\sigma^\prime=\pm}
n_{\mathbf{x},a,\sigma}n_{\mathbf{x},a^\prime,\sigma^\prime}~\!.
\phantom{aaaaaaaaaaa}
\end{eqnarray}
$U$ and $U^\prime$ (both positive) represent here the Coulomb
repulsion of fermions in the same and in different bands, respectively.

Below we consider only a particular version of (\ref{HuMChI}) with
$U=U^\prime$. In this case the notation can be made more concise by
introducing the multi-index $A\equiv(a,\sigma)$:
\begin{eqnarray}
  H =-\sum_{\mathbf{x},\mathbf{y}}\sum_{A}t_{\mathbf{x},\mathbf{y}}~\!
  c^\dagger_{\mathbf{x},A} c_{\mathbf{y},A}
+U\sum_\xgr\sum_{A\ne A'} n_{\xgr,A}n_{\xgr,A'}~\!.\label{HuMChI}
\end{eqnarray}
To estimate the ground state energy of this model one can
introduce the 1- and 2- particle density operators
in full analogy with the definitions (\ref{eqn:1pdo}) and
(\ref{eqn:2pdo}), respectively. The only difference is the
replacement of the spin index $\sigma$ with the multi-index $A$
introduced above. With the help of the density operators
the quantity of interest, the difference
$\Delta E(n)$ defined as in (\ref{eqn:DeltaEdef}),
can be estimated by essentially repeating the steps taken
in the preceding sections.
To this end one introduces the projection Hermitian operators
\begin{eqnarray}
  \hat X_{\mathbf{x}}=\sum_A |\mathbf{x},A\rangle\langle\mathbf{x},A|,
  \phantom{aaa}\sum_{\mathbf{x}} \hat X_{\mathbf{x}}=\hat1_1~\!,\label{XczylinMB}
\end{eqnarray}
which allow to write the interaction term in (\ref{HuMChI})
in the form
\begin{eqnarray}
  \hat V_2=U\sum_{\mathbf{x}}\hat X_{\mathbf{x}}\otimes\hat
  X_{\mathbf{x}}~\!,\label{Uterm}
\end{eqnarray}
analogous to (\ref{V2term}). All considerations used for the single-band
Hubbard Model can now be repeated without any modifications leading to
bounds on $\Delta E$ analogous to (\ref{eqn:FirstBound}).

\subsection{$SU(M)$ symmetric Hubbard Model}

Another version of the Hubbard model to which the method of
\cite{BP} of bounding the difference $\Delta E$ of the gound state
energies can be immediately applied is one in which there are
$M$ kinds (`flavours') of fermions. Its Hamiltonian has the form
\begin{eqnarray}
  H = -\sum_{\mathbf{x},\mathbf{y}}\sum_m t_{\mathbf{x},\mathbf{y}}~\!
  c^\dagger_{\mathbf{x},m} c_{\mathbf{y},m}
+ U\sum_{\mathbf{x}}\sum_{m\neq m^\prime} n_{\mathbf{x},m}n_{\mathbf{x},m^\prime}\nonumber\\
=-\sum_{\mathbf{x},\mathbf{y}}\sum_m t_{\mathbf{x},\mathbf{y}}~\!
c^\dagger_{\mathbf{x},m} c_{\mathbf{y},m}
+\frac{U}{2}\sum_{\mathbf{x}} (n_{\mathbf{x}}^2 - n_{\mathbf{x}})~\!,
\phantom{aaaa}\label{SUM_HuM}
\end{eqnarray}
where $n_{\mathbf{x}}$ is the total number density of particles on the site
$\mathbf{x}$; it can be written in terms of the fermionic creation and
annihilation operators:
$n_{\mathbf{x}}=\sum_{m=1}^M c^\dagger_{\mathbf{x},m} c_{\mathbf{x},m}$. The
operator $n_{\mathbf{x}}$ is invariant under {\em local} $U(M)$ rotations which
mix flavours labeled by different values of $m$. The hopping term,
which involves operators at different sites,
reduces this invariance to the {\em global}
$U(M)$ symmetry only. The $SU(M)$-symmetric Hubbard model is obtained by
stripping off the global $U(1)$ phase factor.
(A certain variant of such a $SU(M)$ Hubbard model, with $M$ having
the interpretation of the number of hyperfine states,  has been
used in \cite{HH} to model cold fermionic gases).
Also in this case, the introduction of the Hermitian projectors 
\begin{eqnarray}
 \hat X_{\mathbf{x}}=\sum_m |\mathbf{x},m\rangle\langle\mathbf{x},m|~\!,
\end{eqnarray}
having all the necessary properties of the projectors (\ref{Xczylin}) and
(\ref{XczylinMB}) allows to rewrite the interaction term of (\ref{SUM_HuM})
in the form analogous to (\ref{V2term}) and (\ref{Uterm}) and, therefore,
the bound on $\Delta E$ established above holds.

\section{Summary}

In this paper we have shown that the method of Bach and Poelchau \cite{BP}
of estimating the difference of the true and the approximate, obtained using
the Hartree-Fock approach, ground state energies of the simplest versions of
the Hubbard models can be extended to more complicated cases of the Hubbard
models with period of the lattice greater than the unit one and to certain
versions of the multi-band Hubbard models.

One possible route to prove the existence of the itinerant ferromagnetism
is to consider the multiband Hubbard model with the Hund interaction, and
obtain a sufficiently tight estimate of $\Delta E$ in this case. The
original motivation for our investigation was the hope that the machinery
eveloped in \cite{BP} can can be extended to obtain the bounds explicitly
depending on the polarization 
for such a realistic version of the Hubbard model of
spin 1/2 fermions on lattices allowing thereby for rigorous 
statements about the existence of ordered phase (at zero temperature)
of such systems. This ambitious goal still remains a challenge
- we don't see how the Hund interactions could be written in the form
similar to (\ref{V2term}), (\ref{Uterm}), which is needed for applying to
it the technique of \cite{BP}. While we still believe
that a lower bound $\Delta E$ can be derived 
in this case as well, we at present we have no idea how to obtain it.
Nevertheless the
method allowed to obtain valuable lower bounds on the difference of the
true ground state energy and the Hartree Fock approximation to it
in some cases which are currently of prime interest but
were not considered in the original work \cite{BP}.


\end{document}

%% file: makra.tex
\newcommand{\bea}{\begin{array}}
\newcommand{\eea}{\end{array}}
\newcommand{\be}{\begin{equation}}
\newcommand{\ee}{\end{equation}}
\newcommand{\ba}{\begin{eqnarray}}
\newcommand{\ea}{\end{eqnarray}}
\newcommand{\baw}{\begin{eqnarray*}}
\newcommand{\eaw}{\end{eqnarray*}}
\newcommand{\bt}{\begin{tabular}}
\newcommand{\et}{\end{tabular}}
\newcommand{\bc}{\begin{center}}
\newcommand{\ec}{\end{center}}
\newcommand{\ben}{\begin{enumerate}}
\newcommand{\een}{\end{enumerate}}
\newcommand{\bi}{\begin{itemize}}
\newcommand{\ei}{\end{itemize}}
\newcommand{\bmpage}{\begin{minipage}}
\newcommand{\empage}{\end{minipage}}

\newcommand{\brak}{\langle}
\newcommand{\kket}{\rangle}

\def \dg{\dagger}
\def \df {{\sf d}}

\def \Del{\Delta}
\def\eps{\epsilon}
\def\veps{\varepsilon}
\def\ga{\gamma}

\def \si {\sigma}

\def\La {\Lambda}

\def\0gr{{\bf 0} }

\def\xgr{{\bf x} }
\def\ygr{{\bf y} }

\def\0gr{{\bf 0} }
\def\bul{$\bullet\;$}

%% file: WCh_060922.bbl
\begin{thebibliography}{99}

\bibitem{BP} V. Bach and J. Poelchau:
  {\em Accuracy of the Hartree-Fock approximation
for the Hubbard model}. {\em J. Math. Phys.} {\bf 38}, 2072-2083 (1997).

\bibitem{Bach_ICMP} V. Bach: {\em Applications of a Correlation Estimate for
Fermions}, in: {\em Proceedings of the XIth International Congress of
Mathematical Physics}, International Press, Boston, 1995, pp. 479-489.

\bibitem{Hubbard} J. Hubbard:
  {\em Electron correlations in narrow energy bands.}
{\em Proc. Roy. Soc. (London)} {\bf A 276}, 238-257 (1963).

\bibitem{Gutzwiller} M. C. Gutzwiller: {\em The effect of correlation on
ferromagnetism of transition metals.} {\em Phys. Rev. Lett.} {\bf 10}, 159-162 (1963). 

\bibitem{Kanamori} J. Kanamori: {\em Electron correlation and ferromagnetism of
transition metals.} {\em Prog. Theor. Phys.} {\bf 30}, 275-289 (1963).

\bibitem{Lieb} E. H. Lieb: {\em The Hubbard model: some rigorous results
and open problems.} In: {\em Proceedings for Advances in Dynamical Systems
and Quantum Physics}, Capri, 1993 (World Scientfic, River Edge, NJ, 1995), pp. 173-193.

\bibitem{BLS}  V. Bach, E. H. Lieb, and J. P. Solovej: {\em Generalized
Hartree-Fock theory and the Hubbard model}. {\em J. Stat. Phys.} {\bf 76},
3-90 (1994).

\bibitem{BLTrav} V. Bach, E. H. Lieb and M. V. Travaglia: {\em Ferromagnetism
of the Hubbard Model at Strong Coupling in the Hartree-Fock Approximation}.
{\em Rev. Math. Phys.} {\bf 18}, 519-543 (2006);  {\sf cond-mat\slash{}0506695}

\bibitem{GrafSolovej} G. M. Graf and J. P. Solovej: {\em A correlation estimate
with applications to quantum systems with Coulomb interactions}. In:
{\em The State of Matter -- A volume dedicated to E. H. Lieb} (World Scientific,
Singapore, 1994); {\em Rev. Math. Phys.} {\bf 6}, 977-997 (1994).

\bibitem{HUMH} T. Hanisch, G. S. Uhrig, and E. M\'{u}ller-Hartmann:
{\em Lattice dependence of saturated ferromagnetism in the Hubbard model}.
{\em Phys. Rev.} {\bf B 56}, 13960 (1997).
  
\bibitem{TasakiFBFM} H. Tasaki: {\em The Hubbard model and origin of ferromagnetism.}
{\em Eur. Phys. J.} {\bf B 64}, 365-372 (2008)

  
\bibitem{Oles} A. M. Ole\'s: {\em Antiferromagnetism and correlation of electrons
  in transition metals. } {\em Phys. Rev.} {\bf B 28}, 327 (1983).

\bibitem{MetallicFM} D. Vollhardt, N. Bl\"umer,
  K. Held and M. Kollar: {\em Metallic Ferromagnetism - An Electronic Correlation
  Phenomenon.}  In: {\em Band-Ferromagnetism. Ground-State and Finite-Temperature
  Phenomena}, Edited by K. Baberschke, M. Donath, W. Nolting, Lecture Notes
  in Physics, vol. 580, p.191 (2001).

\bibitem{HH} C. Honerkamp and W. Hofstetter: {\em BCS pairing in Fermi
systems with N different hyperfine states.} {\em Phys. Rev.} {\bf B 70}, 097521 (2004).
\end{thebibliography}
